\documentclass[pra,aps,showpacs,twocolumn,twoside]{revtex4}
\usepackage{amssymb}
\usepackage{graphicx}
\usepackage{amsmath}
\usepackage{colordvi}

\begin{document}

\title{Optimal entanglement witnesses based on local orthogonal observables}
\author{Cheng-Jie Zhang}
\email{zhangcj@mail.ustc.edu.cn}
\author{Yong-Sheng Zhang}
\email{yshzhang@ustc.edu.cn}
\author{Shun Zhang}
\author{Guang-Can Guo}
\affiliation{Key Laboratory of Quantum Information, University of
Science and Technology of China, Hefei, Anhui 230026, People's
Republic of China}

\begin{abstract}
We show that the entanglement witnesses based on local orthogonal
observables which are introduced in [S. Yu and N.-L. Liu, Phys. Rev.
Lett. \textbf{95}, 150504 (2005)] and [O. G\"uhne, M. Mechler, G.
T\'oth and P. Adam, Phys. Rev. A \textbf{74}, 010301 (R) (2006)] in
linear and nonlinear forms can be optimized, respectively. As
applications, we calculate the optimal nonlinear witnesses of pure
bipartite states and show a lower bound on the I-concurrence of
bipartite higher dimensional systems with our method.
\end{abstract}

\pacs{03.67.Mn, 03.65.Ta, 03.65.Ud}

\maketitle

\section{Introduction}
Entanglement is one of the most fascinating features of quantum
mechanics, which has recently been recognized as a basic resource in
quantum information processing such as teleportation, dense coding
and quantum key distribution \cite{nielsen,Zeilinger}. Thus, it
becomes particularly important to detect and quantify entanglement
\cite{werner}. Despite a great deal of effort in the past years,
lots of things are still unclear to us in this field (see the
reviews \cite{review1,review2,review3} and references therein).
Nevertheless, on the one hand, several sufficient conditions for
detection of entanglement have been found, such as the famous
Peres-Horodecki positive partial transpose (PPT) criterion
\cite{Peres, PPT}, realignment criterion \cite{CCN}, entanglement
witnesses (EWs) \cite{witness1}, local uncertainty relations (LURs)
\cite{LUR1,LUR2}, Bell type inequalities\cite{Bell1,Bell2,Bell3},
etc. PPT criterion is necessary and sufficient for $2\times2$ and
$2\times3$ systems, but only necessary for higher dimensional cases
\cite{PPT}. It is believed that realignment criterion complements
PPT criterion since it can detect many entangled states which PPT
criterion cannot detect. More easier way to detect entanglement
experimentally is using EWs, which have recently been generalized to
nonlinear EWs \cite{witness2,witness3}. On the other hand, a
considerable amount of effort on quantification of entanglement has
also been made. For instance, Wootters has analytically derived a
perfect measure of 2 qubits \cite{concurrence}, which is so-called
$concurrence$. Furthermore, generalized concurrence in bipartite
higher dimensional cases \cite{Uhlmann,I}, such as I-concurrence
\cite{I}, has been pointed out as well. Unfortunately, the
I-concurrence of mixed states is given as a convex roof for all
possible ensemble realization. Therefore, it is generally difficult
to be calculated. Lately, lower bounds on I-concurrence have
attracted much interest \cite{mintert,chen,ph229,ph185}, which are
relatively easier than I-concurrence itself to get.

Recently, Yu and Liu have introduced an entanglement witness [Eq.
(\ref{linear})] based on local orthogonal observables (LOOs) in Ref.
\cite{Yu}. Moreover, G\"{u}hne \textit{et al.} have generalized the
witness to the nonlinear form [Eq. (\ref{nonlinear})] via local
uncertainty relations \cite{nonlinear}. Both of the witnesses have a
common property that each set of LOOs in the witnesses can be
replaced by any other complete set of LOOs, thus one does not know
which set of LOOs is the best one for the witnesses. Actually, the
witnesses using different set of LOOs can obtain distinct results.
For example, the Bell state $(|00\rangle+|11\rangle)/\sqrt{2}$ can
be detected as entangled states by the liner witness under the set
of LOOs: $\{\sigma_{x},\sigma_{y},\sigma_{z},I\}^{A}/\sqrt{2}$,
$\{\sigma_{x},-\sigma_{y},\sigma_{z},I\}^{B}/\sqrt{2}$, but cannot
be detected under the LOOs:
$\{\sigma_{x},\sigma_{y},\sigma_{z},I\}^{A}/\sqrt{2}$,
$\{\sigma_{x},\sigma_{y},\sigma_{z},I\}^{B}/\sqrt{2}$. Therefore, it
is necessary to investigate the optimal case. In this paper, the
optimal witnesses for the linear and nonlinear forms will be
presented. As applications, we will calculate the optimal witnesses
of pure bipartite states and show a lower bound on the I-concurrence
of bipartite higher dimensional systems.

The paper is organized as follows: Sec. II presents the optimal
witnesses of linear and nonlinear forms, which are constructed by
LOOs. In Sec. III we calculate the optimal nonlinear witnesses of
pure bipartite states based on our method. Moreover, we obtain a
lower bound of I-concurrence in bipartite systems. Sec. IV discusses
what happens if the dimensions of the subsystems A and B are not the
same.

\section{Optimal witnesses based on LOOs}
For convenience, we consider a $d\times d$ bipartite system, just as
Refs. \cite{Yu,nonlinear} did ( in Sec. IV we will discuss the
situation when dimensions of subsystems A and B are not the same).
Each subsystem has a complete set of local orthogonal bases
$\{G_{k}^{A}\}$ and $\{G_{k}^{B}\}$, which are so-called LOOs. Such
a basis consists of $d^{2}$ observables and satisfies:
\begin{equation}\label{}
    \mathrm{Tr}(G_{k}^{A}G_{l}^{A})=\mathrm{Tr}(G_{k}^{B}G_{l}^{B})=\delta_{kl}.
\end{equation}
Any other complete set of LOOs relate to the original one by an
orthogonal $d^{2}\times d^{2}$ real matrix, i.e.,
\begin{equation}\label{relate}
    \widetilde{G_{k}^{A}}=\sum_{l}O_{kl}G_{l}^{A},
    \ \widetilde{G_{k}^{B}}=\sum_{l}O'_{kl}G_{l}^{B},
\end{equation}
where $OO^{T}=O^{T}O=O'O'^{T}=O'^{T}O'=I$.

In Ref. \cite{Yu}, a linear witness was introduced as follows (for
convenience, the witness has been written in an equivalent form
introduced in \cite{nonlinear}),
\begin{equation}\label{linear}
    \mathcal{W}=1-\sum_{k}G_{k}^{A}\otimes G_{k}^{B},
\end{equation}
where $\{G_{k}^{A}\}$ and $\{G_{k}^{B}\}$ are arbitrary complete
sets of LOOs for subsystems A and B. Later, Ref. \cite{nonlinear}
provided a nonlinear form,
\begin{equation}\label{nonlinear}
    \mathcal{F}(\rho)=1-\sum_{k}\langle G_{k}^{A}\otimes
    G_{k}^{B}\rangle-\frac{1}{2}\sum_{k}\langle G_{k}^{A}\otimes I-I\otimes
    G_{k}^{B}\rangle^{2}.
\end{equation}
For every separable state $\rho$, it must satisfy that
$\mathrm{Tr}\mathcal{W}\rho\geq0$ and $\mathcal{F}(\rho)\geq0$.
Conversely, if any state violates one of the two inequalities, it is
entangled indeed.

In Refs. \cite{Yu,nonlinear}, there is a little mention involving
that how to choose a set of LOOs so that
$\mathrm{Tr}\mathcal{W}\rho$ or $\mathcal{F}(\rho)$ gets its
minimum, and obviously the minimum means a optimal one, since one
can obtain distinct results by using different sets of LOOs.
Consider the simple example
$|\psi^{+}\rangle=(|00\rangle+|11\rangle)/\sqrt{2}$ introduced in
Sec. I. Under the set of LOOs
$\{\sigma_{x},\sigma_{y},\sigma_{z},I\}^{A}/\sqrt{2}$,
$\{\sigma_{x},\sigma_{y},\sigma_{z},I\}^{B}/\sqrt{2}$,
$\mathrm{Tr}(\mathcal{W}|\psi^{+}\rangle\langle\psi^{+}|)=0$ and
$\mathcal{F}(|\psi^{+}\rangle\langle\psi^{+}|)=0$, with which one
cannot conclude that $|\psi^{+}\rangle$ is entangled. However, under
the set of LOOs
$\{\sigma_{x},\sigma_{y},\sigma_{z},I\}^{A}/\sqrt{2}$,
$\{\sigma_{x},-\sigma_{y},\sigma_{z},I\}^{B}/\sqrt{2}$,
$\mathrm{Tr}(\mathcal{W}|\psi^{+}\rangle\langle\psi^{+}|)=-1$ and
$\mathcal{F}(|\psi^{+}\rangle\langle\psi^{+}|)=-1$. It suggests that
$|\psi^{+}\rangle$ has entanglement. Therefore, it is meaningful to
obtain the minimal one. In the following, we will show that the
minimum is invariant under local unitary (LU) transformations, and
obtain an analytical formula of the minimum.

\textit{Lemma 1.} For a given state $\rho$, the minimum of
$\mathrm{Tr}\mathcal{W}\rho$ [$\mathcal{F}(\rho)$] is LU invariant.

\textit{Proof.}$-$ (Reductio ad absurdum) For a given state $\rho$,
suppose that under the set of LOOs $\{M_{k}^{A}\}$, $\{M_{k}^{B}\}$
$\mathrm{Tr}\mathcal{W}\rho$ [$\mathcal{F}(\rho)$] gets its minimum
$L_{1}$. We operate an arbitrary LU transformation to $\rho$, i.e.,
$\rho'=U_{A}\otimes U_{B}\rho U_{A}^{\dag}\otimes U_{B}^{\dag}$. For
the state $\rho'$, suppose that under the set of LOOs
$\{\widetilde{M_{k}^{A}}\}$, $\{\widetilde{M_{k}^{B}}\}$
$\mathrm{Tr}\mathcal{W}\rho$ [$\mathcal{F}(\rho)$] gets its minimum
$L_{2}$.

Case i. $L_{1}>L_{2}$. For the state $\rho$, under the set of LOOs
$\{U_{A}^{\dag}\widetilde{M_{k}^{A}}U_{A}\}$,
$\{U_{B}^{\dag}\widetilde{M_{k}^{B}}U_{B}\}$,
$\mathrm{Tr}\mathcal{W}\rho$ [$\mathcal{F}(\rho)$] is equal to
$L_{2}$. It is a contradiction to that $L_{1}$ is the minimum of
$\mathrm{Tr}\mathcal{W}\rho$ [$\mathcal{F}(\rho)$].

Case ii. $L_{1}<L_{2}$. For the state $\rho'$, under the set of LOOs
$\{U_{A}M_{k}^{A}U_{A}^{\dag}\}$, $\{U_{B}M_{k}^{B}U_{B}^{\dag}\}$,
$\mathrm{Tr}\mathcal{W}\rho'$ [$\mathcal{F}(\rho')$] is equal to
$L_{1}$. It is a contradiction to that $L_{2}$ is the minimum of
$\mathrm{Tr}\mathcal{W}\rho'$ [$\mathcal{F}(\rho')$].

In a word, if $L_{1}\neq L_{2}$, a contradiction is derived
immediately. Therefore, $L_{1}=L_{2}$ always holds and the minimum
of $\mathrm{Tr}\mathcal{W}\rho$ [$\mathcal{F}(\rho)$] is LU
invariant. \hfill $\square$

\textit{Remark.}$-$ From an experimental point of view, it is
valuable for the minimum to satisfy LU invariant condition, since a
shared spatial reference frame is no longer needed when one makes a
measure of the minimum \cite{G}.

\textit{Theorem 1.} The minimum of $\mathrm{Tr}\mathcal{W}\rho$ is
equal to $1-\sum_{k}\sigma_{k}(\mu)$, where $\sigma_{k}(\mu)$ stands
for the $k$th singular value of real matrix $\mu$ which is defined
as $\mu_{lm}=\mathrm{Tr}(\rho G_{l}^{A}\otimes G_{m}^{B})$.

\textit{Proof.}$-$ Before embarking on our proof, it is worth
noticing that a similar result of Theorem 1 has also been pointed
out in \cite{Yu}. However, for a convenience to understand Theorem
2, we insist on providing a complete proof. For a given state
$\rho$, we choose an arbitrary complete set of LOOs $\{G_{k}^{A}\}$,
$\{G_{k}^{B}\}$. Define that
\begin{equation}\label{}
    \mu_{lm}=\mathrm{Tr}(\rho G_{l}^{A}\otimes G_{m}^{B}),
\end{equation}
and the density matrix can be written as:
\begin{equation}\label{}
    \rho=\sum_{l,m}\mu_{lm}G_{l}^{A}\otimes G_{m}^{B}.
\end{equation}
According to Eq. (\ref{relate}), any other complete set of LOOs
$\{\widetilde{G_{k}^{A}}\}$, $\{\widetilde{G_{k}^{B}}\}$ can be
written as $\widetilde{G_{k}^{A}}=\sum_{l}U_{kl}G_{l}^{A}$,
$\widetilde{G_{k}^{B}}=\sum_{m}V_{km}G_{m}^{B}$, where $U$ and $V$
are $d^{2}\times d^{2}$ real orthogonal matrices, i.e.
$UU^{T}=U^{T}U=VV^{T}=V^{T}V=I$. Therefore,
\begin{eqnarray}
\mathrm{min}\mathrm{Tr}(\mathcal{W}\rho)&=&1-\mathrm{max}\sum_{k}\langle
\widetilde{G_{k}^{A}}\otimes \widetilde{G_{k}^{B}}\rangle\nonumber\\
&=&1-\mathrm{max}\sum_{k}\sum_{lm}U_{kl}V_{km}\langle
G_{l}^{A}\otimes G_{m}^{B}\rangle\nonumber\\
&=&1-\mathrm{max}\sum_{k}\sum_{lm}U_{kl}V_{km}\mu_{lm}\nonumber\\
&=&1-\mathrm{max}\sum_{k}[U\mu V^{T}]_{kk}\nonumber\\
&=&1-\mathrm{max}\mathrm{Tr}(U\mu V^{T}).
\end{eqnarray}
Moreover,
\begin{equation}\label{theorem1}
    \mathrm{max}\mathrm{Tr}(U\mu V^{T})=\mathrm{max}\mathrm{Tr}(\mu V^{T}U)=\sum_{k}\sigma_{k}(\mu),
\end{equation}
where we have used the following theorem \cite{horn}:

\textit{Let $A\in M_{n}$ be a given matrix, and let $A=V\Sigma
W^{\dag}$ be a singular value decomposition of $A$. Then the problem
$max\{Re\ tr AU: U\in M_{n}\ is\  unitary\}$ has the solution
$U=WV^{\dag}$, and the value of the maximum is
$\sigma_{1}(A)+\cdots+\sigma_{n}(A)$, where $\{\sigma_{i}(A)\}$ is
the set of singular values of $A$.}

Notice that $\mu$ is a real matrix and its singular value
decomposition can be written as $\mu=\mathcal{U}^{T}\Sigma
\mathcal{V}$, where $\mathcal{U}$, $\mathcal{V}$ are real orthogonal
matrices and
$\Sigma=diag\{\sigma_{1}(\mu),\sigma_{2}(\mu),\cdots,\sigma_{d^{2}}(\mu)\}$.
When $U=\mathcal{U}$ and $V=\mathcal{V}$, $\mathrm{Tr}(U\mu V^{T})$
gets its maximum $\sum_{k}\sigma_{k}(\mu)$. In other words, under
the new complete set of LOOs $\{\mathcal{G}_{k}^{A}\}$,
$\{\mathcal{G}_{k}^{B}\}$, where
$\mathcal{G}_{k}^{A}=\sum_{l}\mathcal{U}_{kl}G_{l}^{A}$,
$\mathcal{G}_{k}^{B}=\sum_{m}\mathcal{V}_{km}G_{m}^{B}$,
$\mathcal{W}=1-\sum_{k}\mathcal{G}_{k}^{A}\otimes
\mathcal{G}_{k}^{B}$, $\mathrm{Tr}\mathcal{W}\rho$ gets its minimum
$1-\sum_{k}\sigma_{k}(\mu)$. \hfill $\square$

\textit{Remark.}$-$ In fact, it is equivalent to the realignment
criterion when $\mathrm{Tr}\mathcal{W}\rho$ gets its minimum
\cite{Yu}. Note that under the new complete set of LOOs
$\{\mathcal{G}_{k}^{A}\}$, $\{\mathcal{G}_{k}^{B}\}$, the density
matrix can be written in its operator-Schmidt decomposition form
\cite{opeSchde}:
\begin{equation}\label{}
    \rho=\sum_{k}\sigma_{k}(\mu)\mathcal{G}_{k}^{A}\otimes
    \mathcal{G}_{k}^{B}.
\end{equation}
The realignment criterion states that if $\rho$ is separable the sum
of all $\sigma_{k}(\mu)$ is smaller than 1. It is equivalent to
$\mathrm{min} \mathrm{Tr}\mathcal{W}\rho\geq0$. Hence, it is
concluded that any entangled state detected by a witness of Eq.
(\ref{linear}) must violate the realignment criterion.

\textit{Example.}$-$ Let us consider a noisy singlet state
introduced in Ref. \cite{nonlinear}, $\rho=p|\psi_{s}\rangle\langle
\psi_{s}|+(1-p)\rho_{sep}$, where $|\psi_{s}\rangle$ stands for the
singlet state $(|01\rangle-|10\rangle)/\sqrt{2}$ and the separable
noise is
$\rho_{sep}=2/3|00\rangle\langle00|+1/3|01\rangle\langle01|$.
Actually, the state is entangled for any $p>0$ \cite{nonlinear}.
Under the complete set of LOOs
$\{-\sigma_{x},-\sigma_{y},-\sigma_{z},I\}^{A}/\sqrt{2}$,
$\{\sigma_{x},\sigma_{y},\sigma_{z},I\}^{B}/\sqrt{2}$, the witness
of Eq. (\ref{linear}) can detect the entanglement for all $p>0.4$.
However, the optimal witness using Theorem 1 can detect the
entanglement for all $p>0.292$, which is equivalent to the
realignment criterion.

\textit{Theorem 2.} The minimum of $\mathcal{F}(\rho)$ is equal to
$1-\sum_{k}\sigma_{k}(\tau)-(\mathrm{Tr}\rho_{A}^{2}+\mathrm{Tr}\rho_{B}^{2})/2$,
where $\sigma_{k}(\tau)$ stands for the $k$th singular value of
matrix $\tau$ defined as $\tau_{lm}=\langle G_{l}^{A}\otimes
G_{m}^{B}\rangle-\langle G_{l}^{A}\otimes I\rangle\langle I\otimes
G_{m}^{B}\rangle$.

\textit{Proof.}$-$ For a given state $\rho$, we choose an arbitrary
complete sets of LOOs $\{G_{k}^{A}\}, \{G_{k}^{B}\}$, and calculate
the real matrix $\tau$ according to the definition:
\begin{equation}\label{}
    \tau_{lm}=\langle G_{l}^{A}\otimes G_{m}^{B}\rangle-\langle
G_{l}^{A}\otimes I\rangle\langle I\otimes G_{m}^{B}\rangle.
\end{equation}
Similarly to Theorem 1, any other complete set of LOOs
$\{\widetilde{G_{k}^{A}}\}$, $\{\widetilde{G_{k}^{B}}\}$ can be
written as $\widetilde{G_{k}^{A}}=\sum_{l}U_{kl}G_{l}^{A}$,
$\widetilde{G_{k}^{B}}=\sum_{m}V_{km}G_{m}^{B}$, where $U$ and $V$
are $d^{2}\times d^{2}$ real orthogonal matrices, i.e.
$UU^{T}=U^{T}U=VV^{T}=V^{T}V=I$. Therefore,

\begin{eqnarray}
  &&\mathrm{min} [1-\sum_{k}\langle \widetilde{G_{k}^{A}}\otimes \widetilde{G_{k}^{B}}\rangle-\frac{1}{2}\sum_{k}\langle
  \widetilde{G_{k}^{A}}\otimes I-I\otimes
  \widetilde{G_{k}^{B}}\rangle^{2}]\nonumber\\
   &=& 1- \mathrm{max} [\sum_{k}\langle \widetilde{G_{k}^{A}}\otimes \widetilde{G_{k}^{B}}\rangle+\frac{1}{2}\sum_{k}\langle \widetilde{G_{k}^{A}}\otimes I-I\otimes
   \widetilde{G_{k}^{B}}\rangle^{2}].\nonumber
\end{eqnarray}
Moreover,
\begin{eqnarray}
&&\sum_{k}\langle \widetilde{G_{k}^{A}}\otimes I-I\otimes
   \widetilde{G_{k}^{B}}\rangle^{2}\nonumber\\
   &=&\sum_{k}[\langle \widetilde{G_{k}^{A}}\otimes I\rangle^{2}+\langle I\otimes \widetilde{G_{k}^{B}}\rangle^{2}-2\langle \widetilde{G_{k}^{A}}\otimes I\rangle\langle I\otimes
   \widetilde{G_{k}^{B}}\rangle],\nonumber
\end{eqnarray}
where $\sum_{k}\langle \widetilde{G_{k}^{A}}\otimes I\rangle^{2}$
and $\sum_{k}\langle I\otimes \widetilde{G_{k}^{B}}\rangle^{2}$ are
invariant under LOOs transformations, i.e.,
\begin{eqnarray}
  \sum_{k}\langle \widetilde{G_{k}^{A}}\otimes I\rangle^{2} &=& \sum_{k}\sum_{ll'}U_{kl}U_{kl'}\langle G_{l}^{A}\otimes I\rangle\langle G_{l'}^{A}\otimes I\rangle\nonumber\\
   &=& \sum_{ll'}[U^{T}U]_{ll'}\langle G_{l}^{A}\otimes I\rangle\langle G_{l'}^{A}\otimes
   I\rangle\nonumber\\
   &=& \sum_{l}\langle G_{l}^{A}\otimes
   I\rangle^{2}\nonumber\\
   &=& \mathrm{Tr}\rho_{A}^{2},\nonumber
\end{eqnarray}
where $\rho_{A}$ is the reduced density matrix after tracing over
subsystem B. Without loss of generality, substituting Eq.
(\ref{lambda}) into $\sum_{l}\langle G_{l}^{A}\otimes I\rangle^{2}$,
one can obtain the final result $\mathrm{Tr}\rho_{A}^{2}$.
Similarly, $\sum_{k}\langle
I\otimes\widetilde{G_{k}^{B}}\rangle^{2}=\sum_{l}\langle I\otimes
G_{l}^{B}\rangle^{2}=\mathrm{Tr}\rho_{B}^{2}$ holds.
\begin{eqnarray}
G_{k}^{A}=\left\{
\begin{array}{ll}
\frac{1}{\sqrt{2}}(|m\rangle\langle n|+|n\rangle\langle m|) & 1\leq
m<n\leq d,\\
\frac{1}{\sqrt{2}}(i|m\rangle\langle n|-i|n\rangle\langle m|) &
1\leq m<n\leq d,\\
|m\rangle\langle m| & 1\leq m\leq d.
\end{array}\right.\label{lambda}
\end{eqnarray}
\begin{eqnarray}
G_{k}^{B}=(G_{k}^{A})^{T},\label{lambda2}
\end{eqnarray}
where $\{|m\rangle_{A}\}$ and $\{|m\rangle_{B}\}$ are the standard
complete bases. Thus,
\begin{eqnarray}
&&\mathrm{max} [\sum_{k}\langle \widetilde{G_{k}^{A}}\otimes
\widetilde{G_{k}^{B}}\rangle+\frac{1}{2}\sum_{k}\langle
\widetilde{G_{k}^{A}}\otimes I-I\otimes
   \widetilde{G_{k}^{B}}\rangle^{2}]\nonumber\\
&=&\frac{1}{2}\sum_{k}[\langle \widetilde{G_{k}^{A}}\otimes
I\rangle^{2}+\langle I\otimes
\widetilde{G_{k}^{B}}\rangle^{2}]\nonumber\\
&&+ \mathrm{max} [\sum_{k}(\langle \widetilde{G_{k}^{A}}\otimes
\widetilde{G_{k}^{B}}\rangle-\langle \widetilde{G_{k}^{A}}\otimes
I\rangle\langle I\otimes\widetilde{G_{k}^{B}}\rangle)]\nonumber\\
&=& \frac{1}{2}(\mathrm{Tr}\rho_{A}^{2}+\mathrm{Tr}\rho_{B}^{2})+
\mathrm{max}
\sum_{k}\sum_{lm}U_{kl}V_{km}\tau_{lm}\nonumber\\
&=& \frac{1}{2}(\mathrm{Tr}\rho_{A}^{2}+\mathrm{Tr}\rho_{B}^{2})+
\mathrm{max} \sum_{k}[U\tau
V^{T}]_{kk}\nonumber\\
&=& \frac{1}{2}(\mathrm{Tr}\rho_{A}^{2}+\mathrm{Tr}\rho_{B}^{2})+
\sum_{k}\sigma_{k}(\tau).\label{theorem2}
\end{eqnarray}
In other words, $\mathrm{min}
\mathcal{F}(\rho)=1-\sum_{k}\sigma_{k}(\tau)-(\mathrm{Tr}\rho_{A}^{2}+\mathrm{Tr}\rho_{B}^{2})/2$.
\hfill $\square$

\textit{Example.}$-$ Bennett {\it et al.} introduced a $3\times3$
bound entangled state constructed from unextendible product bases in
Ref. \cite{UPB}:
\begin{eqnarray}
|\psi_{0}\rangle=\frac{1}{\sqrt{2}}|0\rangle(|0\rangle-|1\rangle),
\ |\psi_{1}\rangle=\frac{1}{\sqrt{2}}(|0\rangle-|1\rangle)|2\rangle,\nonumber\\
|\psi_{2}\rangle=\frac{1}{\sqrt{2}}|2\rangle(|1\rangle-|2\rangle),
\ |\psi_{3}\rangle=\frac{1}{\sqrt{2}}(|1\rangle-|2\rangle)|0\rangle,\nonumber\\
|\psi_{4}\rangle=\frac{1}{3}(|0\rangle+|1\rangle+|2\rangle)(|0\rangle+|1\rangle+|2\rangle),\nonumber
\end{eqnarray}
\begin{equation}\label{upb}
    \rho=\frac{1}{4}(I-\sum_{i=0}^{4}|\psi_{i}\rangle\langle\psi_{i}|).
\end{equation}

Let us consider a mixture of this state with white noise,
\begin{equation}\label{upbmix}
    \rho(p)=p\rho+(1-p)\frac{I}{9}.
\end{equation}
Using the realignment criterion, one finds that the state $\rho(p)$
still has entanglement when $p>0.8897$. In Ref. \cite{nonlinear}, it
is found that the state $\rho(p)$ must be entangled for
$p>p_{lur}=0.8885$ using the nonlinear witness Eq. (\ref{nonlinear})
(but not the optimal one). According to Theorem 2, one can obtain an
optimal witness of Eq. (\ref{nonlinear}) and find that when
$p>p_{opt}=0.8822$ the state is still entangled. Obviously, the
optimal witness is stronger than the one in Ref. \cite{nonlinear}.
In addition, in Sec. III we will present a lower bound on
I-concurrence for the state based on Theorem 2 (see Fig. 1). From
the figure, it is worth noticing that the bound is positive when
$p>p_{opt}=0.8822$.

\section{Applications}
In this section, the optimal nonlinear witnesses of pure bipartite
states will be obtained using Theorem 2. Moreover, we will show a
lower bound on the I-concurrence of bipartite systems by means of
our method. Before embarking on our investigation, we first define
that $\mathcal{L}=\frac{1}{2}\sum_{k}\langle G_{k}^{A}\otimes
I-I\otimes G_{k}^{B}\rangle^{2}+\sum_{k}\langle G_{k}^{A}\otimes
G_{k}^{B}\rangle$, and obviously
$\mathcal{L}_{max}=\sum_{k}\sigma_{k}(\tau)+(\mathrm{Tr}\rho_{A}^{2}+\mathrm{Tr}\rho_{B}^{2})/2$
according to Theorem 2.

\subsection{Optimal witnesses of bipartite pure states}
Let us calculate $\mathcal{L}_{max}$ of a bipartite pure state
$|\psi\rangle$ with its Schmidt decomposition
$|\psi\rangle=\sum_{i}\sqrt{\mu_{i}}|ii\rangle$.

Since Schmidt decomposition of a pure state is a LU transformation,
$\mathcal{L}_{max}(|\psi\rangle)$ remains invariant after the
transformation according to Lemma 1. Therefore, we can directly use
the Schmidt decomposition form for convenience. We choose a complete
set of LOOs Eq. (\ref{lambda}) and Eq. (\ref{lambda2}) for A and B
subsystems, respectively (obviously any other complete set of LOOs
can be chosen and it does not affect the final result).

According to Theorem 2,
\begin{eqnarray}
  \tau_{lm}&=&\langle
G_{l}^{A}\otimes G_{m}^{B}\rangle-\langle G_{l}^{A}\otimes
I\rangle\langle I\otimes G_{m}^{B}\rangle\nonumber \\
   &=& [D\oplus D\oplus T]_{lm},
\end{eqnarray}
where
$D=diag\{\sqrt{\mu_{1}\mu_{2}},\cdots,\sqrt{\mu_{m}\mu_{n}}\cdots,\sqrt{\mu_{d-1}\mu_{d}}\}$
and
\begin{eqnarray}
T=\left(\begin{array}{cccc} \mu_{1}-\mu_{1}^{2} & -\mu_{1}\mu_{2}&
\cdots & -\mu_{1}\mu_{d}\\
-\mu_{1}\mu_{2} & \mu_{2}-\mu_{2}^{2} & \cdots & -\mu_{2}\mu_{d}\\
\vdots &\vdots & \ddots &\vdots \\
-\mu_{1}\mu_{d} & -\mu_{2}\mu_{d} & \cdots & \mu_{d}-\mu_{d}^{2}
\end{array}\right).
\end{eqnarray}
Therefore,
\begin{eqnarray}
&&\sum_{k}\sigma_{k}(\tau)=2\sum_{m<n}\sqrt{\mu_{m}\mu_{n}}+2\sum_{m<n}\mu_{m}\mu_{n},\\
&&\frac{1}{2}(\mathrm{Tr}\rho_{A}^{2}+\mathrm{Tr}\rho_{B}^{2})=\sum_{i}\mu_{i}^{2},\\
&&\mathcal{L}_{max}(|\psi\rangle)=(\sum_{i}\sqrt{\mu_{i}})^{2}.\label{lmax}
\end{eqnarray}

Note that Eq. (\ref{lmax}) has also been derived with another
totally different method in Ref. \cite{ph229}, and it completely
accords with our result. Compared with the method in Ref.
\cite{ph229}, Theorem 2 in this paper is more general, i.e., it
suits not only bipartite pure states but also any bipartite mixed
state.

\subsection{Lower bound on the I-concurrence}
I-concurrence of a bipartite pure state is given by
$C(|\psi\rangle)=\sqrt{2(1-\mathrm{Tr}\rho_{A}^{2})}$, where the
reduced density matrix $\rho_{A}$ is obtained by tracing over the
subsystem B. It can be extended to mixed states $\rho$ by the convex
roof,
\begin{equation}\label{}
    C(\rho)=\inf_{\{p_{i},|\psi_{i}\rangle\}} \sum_{i}p_{i}C(|\psi_{i}\rangle),
    \   \rho=\sum_{i}p_{i}|\psi_{i}\rangle\langle\psi_{i}|,
\end{equation}
for all possible decomposition into pure states, where $p_{i}\geq0$
and $\sum_{i}p_{i}=1$.

Several bounds have already been derived
\cite{mintert,chen,ph229,ph185}, e.g., an analytical lower bound
based on PPT criterion and the realignment criterion has been shown
by Chen \textit{et al.},
\begin{equation}\label{}
    C(\rho)\geq \sqrt{\frac{2}{m(m-1)}}(\mathrm{max}
    (\|\rho^{T_{A}}\|,\|\mathcal{R}(\rho)\|)-1),
\end{equation}
where $T_{A}$, $\mathcal{R}$ and $\|\cdot\|$ stand for partial
transpose, realignment and the trace norm (i.e. the sum of the
singular values), respectively. In Ref. \cite{ph229}, another bound
based on LOOs has been obtained, which has used Eq. (\ref{lmax}) and
the fact that
$\sum_{i}p_{i}\mathcal{L}_{max}(|\psi_{i}\rangle)\geq\sum_{i}p_{i}\mathcal{L}(|\psi_{i}\rangle)\geq\mathcal{L}(\sum_{i}p_{i}|\psi_{i}\rangle\langle\psi_{i}|)$,
(for convenience, the lower bound has been rewritten in an
equivalent form)
\begin{equation}\label{C}
    C(\rho)\geq \sqrt{\frac{2}{m(m-1)}}(\mathcal{L}-1).
\end{equation}
Notice that Eq. (\ref{C}) holds for arbitrary set of LOOs, including
the optimal one. Therefore, a tighter form of Eq. (\ref{C}) can be
obtained according to Theorem 2,
\begin{equation}\label{Cl}
    C(\rho)\geq \sqrt{\frac{2}{m(m-1)}}(\mathcal{L}_{max}-1),
\end{equation}
where
$\mathcal{L}_{max}=\sum_{k}\sigma_{k}(\tau)+(\mathrm{Tr}\rho_{A}^{2}+\mathrm{Tr}\rho_{B}^{2})/2$.
Since the entanglement criteria based on LURs are strictly stronger
than the realignment criterion \cite{nonlinear}, the following
inequality can be concluded.
\begin{equation}\label{}
    C(\rho)\geq \sqrt{\frac{2}{m(m-1)}}(\mathrm{max}
    (\|\rho^{T_{A}}\|,\mathcal{L}_{max}(\rho))-1).
\end{equation}
For example, reconsider the bound entangled state Eq. (\ref{upb}).
Because it belongs to PPT entangled state, the lower bound based on
PPT criterion is unhelpful. One can obtain that $C(\rho)\geq0.050$
via the realignment criterion, and $C(\rho)\geq0.052$ has been
gotten in Ref. \cite{ph229} by using Eq. (\ref{C}). In fact,
$\mathcal{L}_{max}(\rho)$ can be directly calculated, and it
suggests that $C(\rho)\geq0.055$ via Eq. (\ref{Cl}), which is better
than the one in Ref. \cite{ph229}. Furthermore, one can consider the
bound entangled state with white noise, i.e. Eq. (\ref{upbmix}). The
lower bounds of I-concurrence for $\rho(p)$ have been shown in Fig.
1. Therefore, the lower bound based on $\mathcal{L}_{max}$ has been
strictly improved compared with the one based on the realignment
criterion and provided a tighter form of Eq. (\ref{C}).

\begin{figure}
\begin{center}
\includegraphics[scale=0.4]{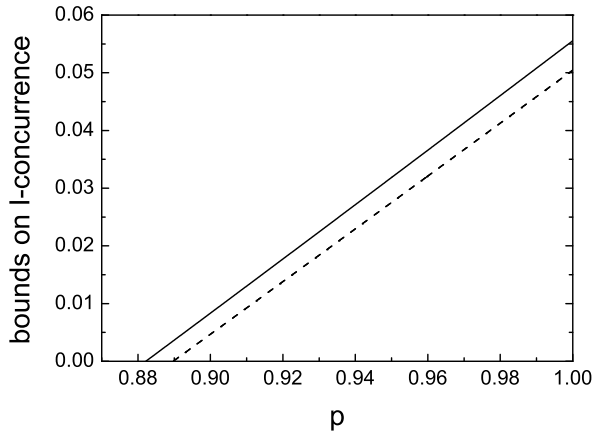}
\caption{Two lower bounds of I-concurrence for the state $\rho(p)$.
One is the lower bound based on realignment criterion (dashed line),
the other is obtained from $\mathcal{L}_{max}$ (solid line).}
\end{center}
\end{figure}

\section{Discussion and conclusion}
During the last two sections, we consider a simple situation: the
$d\times d$ bipartite system for convenience. However, if the
dimensions of the Hilbert spaces $\mathcal{H}_{A}$ and
$\mathcal{H}_{B}$ are not the same, what will happen? Actually, it
does not affect any one of the conclusions in Sec. II and Sec. III.

Without loss of generality, suppose that
$m=\mathrm{dim}(\mathcal{H}_{A})<n=\mathrm{dim}(\mathcal{H}_{B})$.
There are $m^{2}$ elements in a complete set of LOOs
$\{G_{k}^{A}\}$, and $n^{2}$ elements in $\{G_{k}^{B}\}$. Therefore,
we need to reconsider Eq. (\ref{theorem1}) and Eq. (\ref{theorem2})
in Theorem 1 and Theorem 2, respectively.
\begin{eqnarray}
\mathrm{max}\sum_{k}\sum_{lm}U_{kl}V_{km}\mu_{lm}=\mathrm{max}\mathrm{Tr}(U\mu
V^{T}),\\
\mathrm{max}\sum_{k}\sum_{lm}U_{kl}V_{km}\tau_{lm}=\mathrm{max}\mathrm{Tr}(U\tau
V^{T}),\label{th2}
\end{eqnarray}
where $U$ is an $m^{2}\times m^{2}$ real orthogonal matrix; $\mu$
and $\tau$ are $m^{2}\times n^{2}$ real matrices; $V$ belongs to
$n^{2}\times n^{2}$ real orthogonal matrices. The two equations have
the same form, so we just need to consider Eq. (\ref{th2}) for
instance.

As Ref. \cite{nonlinear} did, one can define that $G_{k}^{A}=0$ for
$k=m^{2}+1, \cdots, n^{2}$. Thus, the matrix $\tau$ is changed into
an $n^{2}\times n^{2}$ real matrix, i.e.,
\begin{equation}\label{}
   \tau'=\left(\begin{array}{c}
    \tau\\
    0
   \end{array}\right),
\end{equation}
where $0$ stands for an $(n^{2}-m^{2})\times n^{2}$ matrix with
every element being equal to $0$.

Define that $U'=U\oplus I$, where $I$ is an
$(n^{2}-m^{2})\times(n^{2}-m^{2})$ identity matrix. It is easy to
see that $U'$ is an $n^{2}\times n^{2}$ real orthogonal matrix since
$U$ belongs to $m^{2}\times m^{2}$ real orthogonal matrices.

Notice that ($l\equiv n^{2}-m^{2}$)
\begin{equation}\label{}
    \left(\begin{array}{cc}
    U_{m^{2}\times m^{2}} & 0\\
    0 & I_{l\times l}
   \end{array}\right)
   \left(\begin{array}{c}
    \tau_{m^{2}\times n^{2}}\\
    0_{l\times n^{2}}
   \end{array}\right)
   \left(\begin{array}{c}
    V^{T}_{n^{2}\times n^{2}}
   \end{array}\right)=
   \left(\begin{array}{c}
    [U\tau V^{T}]_{m^{2}\times n^{2}}\\
    0_{l\times n^{2}}
   \end{array}\right),\nonumber
\end{equation}
which means that $\mathrm{Tr}[U'\tau' V^{T}]=\mathrm{Tr}[U\tau
V^{T}]$. Therefore,
\begin{eqnarray}
\mathrm{max} \mathrm{Tr}[U\tau V^{T}]&=&\mathrm{max}
\mathrm{Tr}[U'\tau' V^{T}]\nonumber\\
&=& \mathrm{max} \mathrm{Tr}[\tau' V^{T}U']\nonumber\\
&=& \sum_{k}\sigma_{k}(\tau').\label{tr1}
\end{eqnarray}
Since $\tau'\tau'^{T}=[\tau\tau^{T}]\oplus 0_{l\times l}$,
$\tau'\tau'^{T}$ and $\tau\tau^{T}$ have the same nonzero
eigenvalues. Hence,
\begin{equation}\label{singular}
    \sum_{k}\sigma_{k}(\tau')=\sum_{k}\sigma_{k}(\tau).
\end{equation}

Consequently, Eq. (\ref{tr1}) and Eq. (\ref{singular}) suggest that
Theorem 1 and Theorem 2 still hold even if the dimensions of
subsystems A and B are not the same, and the applications in Sec.
III which have used the Theorem 2 can also be extended to this case.

In conclusion, we have optimized the linear and the nonlinear
entanglement witnesses based on local orthogonal observables, which
are introduced by Yu, Liu and G\"{u}hne \textit{et al.}
respectively, and several examples have been given as well.
Moreover, we have obtained the optimal witnesses based on LOOs in
pure bipartite systems and a lower bound on the I-concurrence of
bipartite systems as applications of our method. In fact, Theorem 2
presents a separability criterion with Ky Fan norm of $\tau$, the
covariance term defined in \cite{G}. Similarly, another separability
criterion with Ky Fan norm of correlation matrix has been shown in
\cite{bloch}. It is worth investigating deeper relation between this
two criterions. In addition, the `optimal' in this paper is in the
sense of choosing the best complete set of LOOs such that the
witness gets its minimum, which has little relation with traditional
optimal EWs \cite{optimal}.

\textit{Note added.} Recently a similar result has been shown in
\cite{ph282}, which is based on covariance matrix criterion.
Interestingly, Proposition 3 in \cite{ph282} can be optimized to a
similar form as Theorem 2 in this paper.

\section*{ACKNOWLEDGMENTS}
This work was funded by the National Fundamental Research Program
(2006CB921900), the National Natural Science Foundation of China
(10674127, 60121503), the Innovation Funds from the Chinese Academy
of Sciences, and Program for New Century Excellent Talents in
University.

\end{document}